\documentclass[aps,prl,twocolumn, groupedaddress]{revtex4-1}
\usepackage{graphicx}
\usepackage{amsmath}
\usepackage{subfigure}

\begin{document}
\title{Manipulation of a continuous beam of molecules by light pulses}
\author{Paul Venn}
\author{Hendrik Ulbricht}
\email[]{h.ulbricht@soton.ac.uk}
\affiliation{Physics and Astronomy, University of Southampton,
Highfield, Southampton, SO17 1BJ, UK}
\date{\today}

\begin{abstract}
We experimentally observe the action of multiple light pulses on the transverse motion of a continuous beam of fullerenes. The light potential is
generated by non-resonant ultra-short laser pulses in
perpendicular spatial overlap with the molecule beam. We observe
a small but clear enhancement of the number of molecules in the
center fraction of the molecular beam. Relatively low light
intensity and short laser pulse duration prevent the molecule
from fragmentation and ionization. Experimental results are
confirmed by  Monte Carlo trajectory simulations. 
\end{abstract}


\maketitle
It is known from both theory~\cite{Seidman96} and
experiment~\cite{Stapelfeldt97,Sakai98,zhao2000,fulton2004optical}
that when a neutral molecule enters the focus of a time-varying electric
field a dipole force is acting on the center of mass motion of the particle. The same effect is used for optical tweezing of micro-meter sized particles and biological cells. The dipole potential $U$ is related to the dynamic (frequency dependent) polarizability of the molecule, $\alpha$,
and the space and time dependent distribution of the intensity 
of a light field $E^{2}$: $U\left(x,y,z,t\right)=-\frac{1}{4}\alpha E^{2}\left(x,y,z,t\right). $

The dipole force, $F=-\nabla U$, is proportional to the gradient of the laser intensity. Assuming a Gaussian laser profile, the velocity change of the molecules in the $y$-direction (see Fig. 2) is obtained by integrating the force over light-matter interaction time, $\Delta v_{y}=\frac{1}{m}\int_{-\infty}^{\infty} \!
F_{y}(t)\,\mathrm{d}t$, yielding:

\begin{equation}
\Delta v_{y}=-4y\sqrt\frac{{\pi}}{{2}}\frac{U}{mv_{x}w_{0}}\frac{1}{\sqrt{1+2\ln2\left(
\frac{w_{0}}{v_{x}\tau}\right)^{2}}}\exp\left(\frac{-2y^{2}}{{w_{0}}^{2}}
\right),
\label{eq:vel}
\end{equation}
where $x=v_{x}t$ describes the longitudinal motion of the molecule and $\tau$ is the light-molecule interaction time.
\begin{figure}[t]
\hspace*{0.1cm}
\subfigure(a){\includegraphics[width=0.40\textwidth]{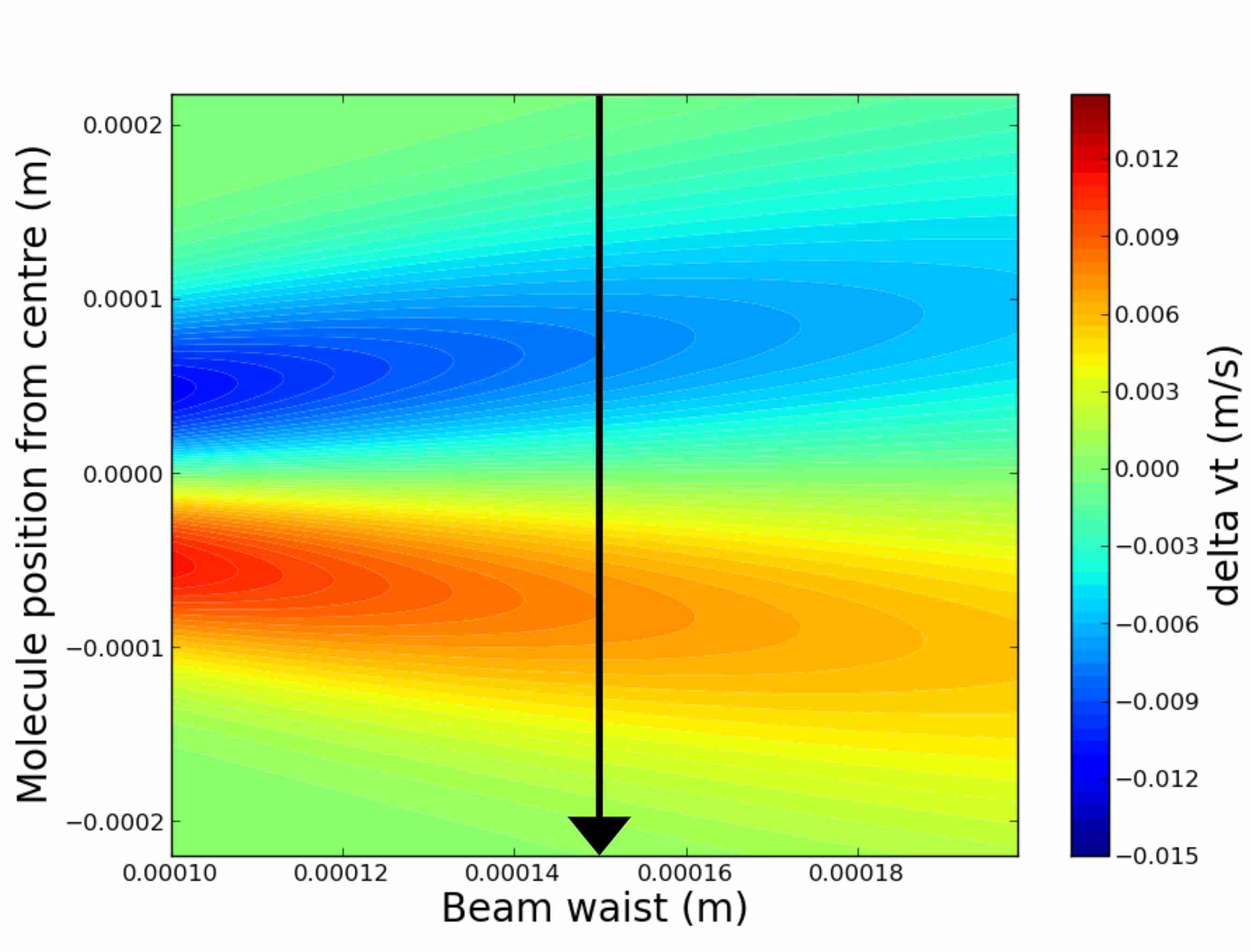}}
\subfigure(b){\includegraphics[width=0.40\textwidth]{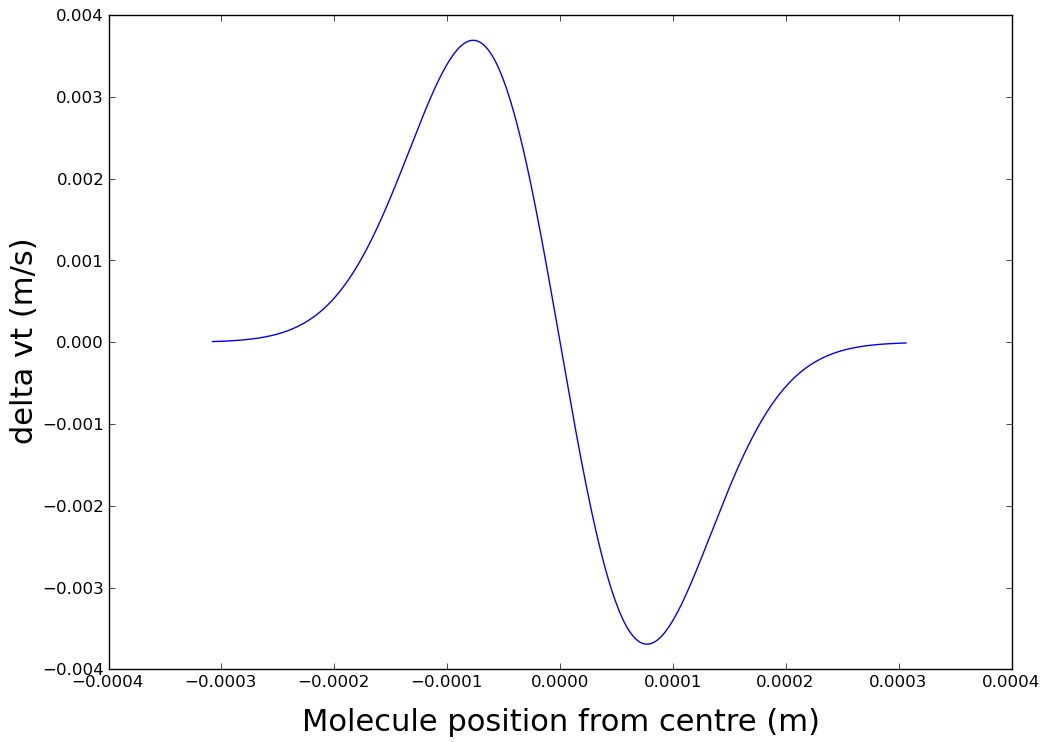}}
\caption{(a) Model showing the total change in velocity of a
test molecule ($C_{60}$) traveling with initially zero
transverse velocity (perfect collimation) through a focused laser for different beam
waists and positions along the Gaussian profile of the laser.
(b) total velocity change for a beam waist of 154 ${\mu}$m, cut
through (a) along arrow. In both simulations the following
parameters were used $m_{C60}$=720 amu, $v_x$=180 m/s, pulse
length=100 fs, ${\alpha}_{C60}$=90 $\AA^{3}$, r=76 MHz,
$P_{peak}$=60 kW.}
\label{fig:models}
\end{figure}
Earlier experiments using dipole force observed the change in velocity for a pulsed beam of small molecules interacting with an individual tightly focused laser pulse of diameter ~10 $\mu$m~\cite{zhao2000}. In contrast we will measure the transverse effect by its net increase in molecular beam flux at a certain spatial area at the detector. 

\begin{figure}[t]
\centering
\includegraphics[width=0.40\textwidth]{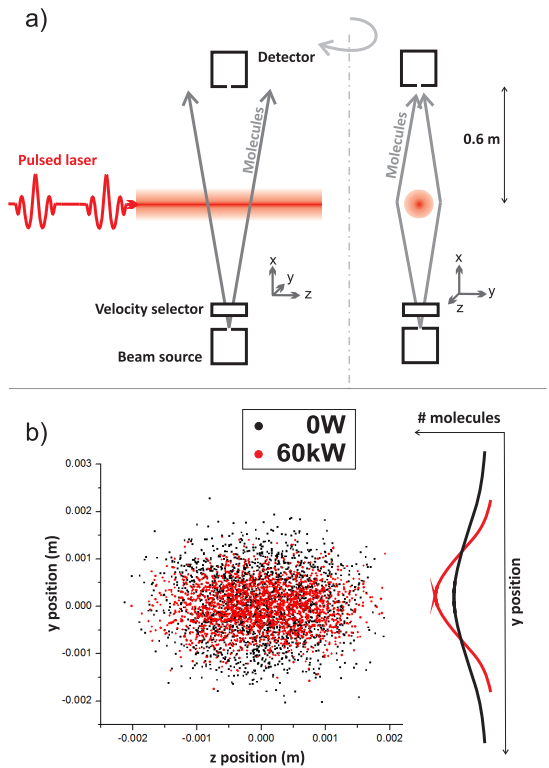}
\caption{(a) Schematics of the experimental setup and the
geometry of the focusing effect. This affects only one spatial
direction of the molecule beam due to the light intensity
gradient - an elliptic lens. Light focusing in z-direction is
too weak to have an effect. (b) Molecule as counted in the
detector plane by Monte Carlo trajectory simulations showing the
qualitative effect of the lensing with laser (red points)
compared to detected molecules without laser (black points) for
the same parameter as for the simulations of the dipole potential
in Fig.~\ref{fig:models}. The energy is per laser pulse.}
\label{fig:monte carlo}
\end{figure}

First, we model the dipole force effect on the motion of neutral molecules for a quasi-continuous laser beam of increasing the laser waist $w_0$ where a test particle is propagating through the potential energy landscape of focused light (Eqn.~\ref{eq:vel}). The calculated total change in transverse velocity ($\Delta v_{t}$) for a molecule passing through the laser spot of different waist is shown in Fig.~$\ref{fig:models}$a). A single dispersion profile for $w_0$=154 ${\mu}m$ is shown in Fig.~$\ref{fig:models}$b). As expected, increasing
the beam waist at a given laser power leads to a reduction in the transverse velocity effect. second, from trajectory simulation by randomly sampling of starting conditions for position and transverse velocity (Monte Carlo) we model the effect of multiple pulses acting on individual molecule trajectories. Molecule distributions simulated with and without laser interactions are shown in Fig.~\ref{fig:monte carlo} b) indicating a clear squeezing of the spatial molecule distribution in y-direction for the case with laser 'on'.  

Experiments have been performed with $C_{60}$ fullerene beams formed by sublimation in an oven (Sigma Aldrich, 99.9$\%$ purity). The longitudinal velocity $v_{x}$ was selected to be
180 m/s with a longitudinal spread of $\Delta v/v = {\pm}2.2\%$ (FWHM)~\cite{Szewc10}. The molecular beam is collimated by a 1 mm aperture (collimation is about 1 mrad) before it is crossed with a pulsed laser (Coherent MIRA, pulse duration 100 fs, peak power 10 nJ, wavelength 800 nm) aligned along the z-axis. The laser beam was focused by a f = 100 cm lens to have a
waist of about 100 $\mu$m at the light-molecule crossing.
The vacuum chamber was kept at a pressure of
$1{\times}10^{-8}$ mbar. See for setup Fig.~$\ref{fig:monte carlo}$a). Molecules are detected by a Quadrupole Mass Spectrometer (Extrel) aligned in the x-axis, at a distance of 0.6 m after the light-molecule crossing. Spatial cross sections of the molecular beam were detected by moving the detector position with respect to the molecule beam or using sub-mm apertures and slits
aligned in the z- and y-axis in front of the detector. 

\begin{figure}[t]
\hspace*{0.1cm}
\subfigure(a){\includegraphics[width=0.40\textwidth]{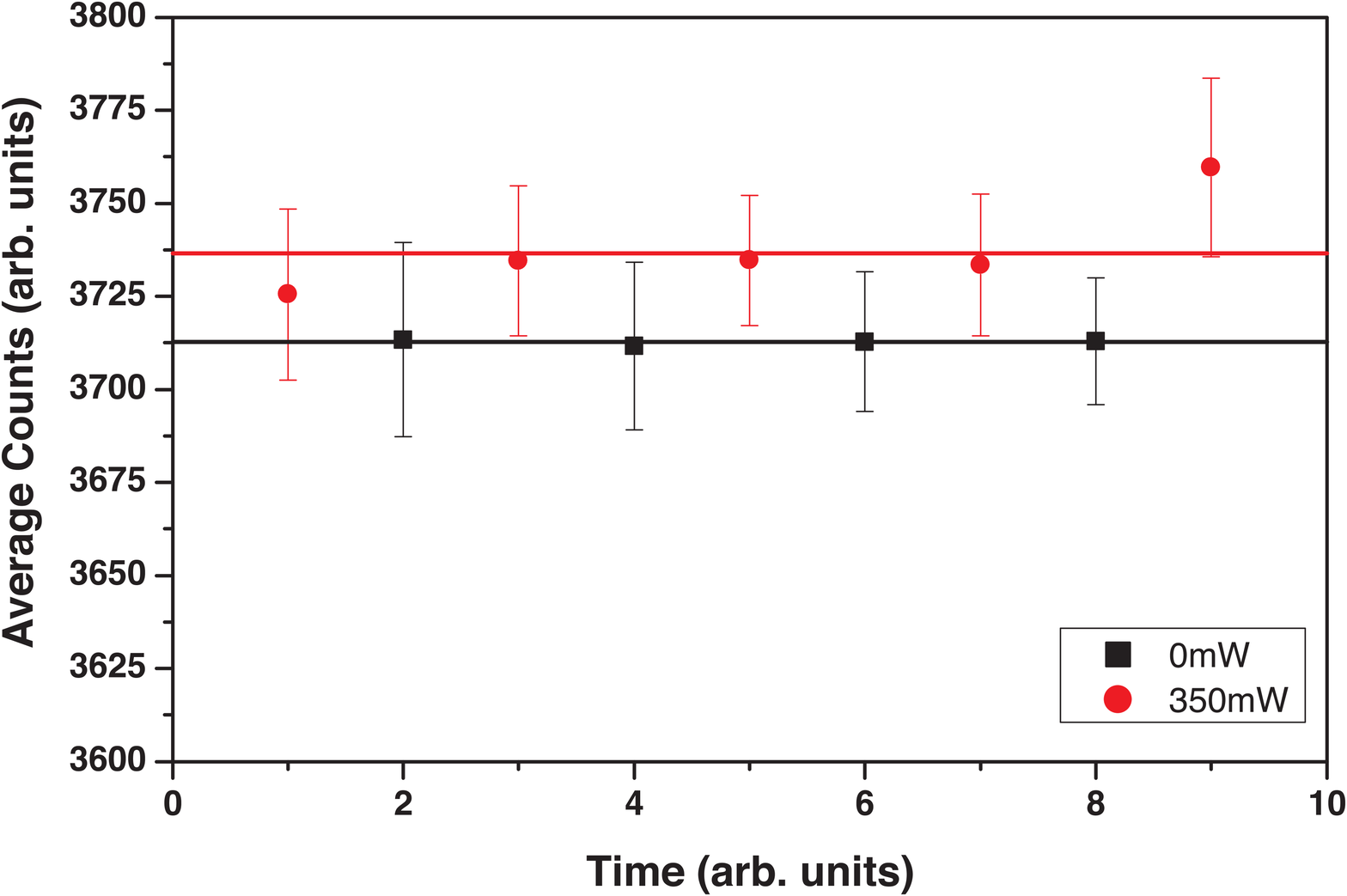}}
\subfigure(b){\includegraphics[width=0.40\textwidth]{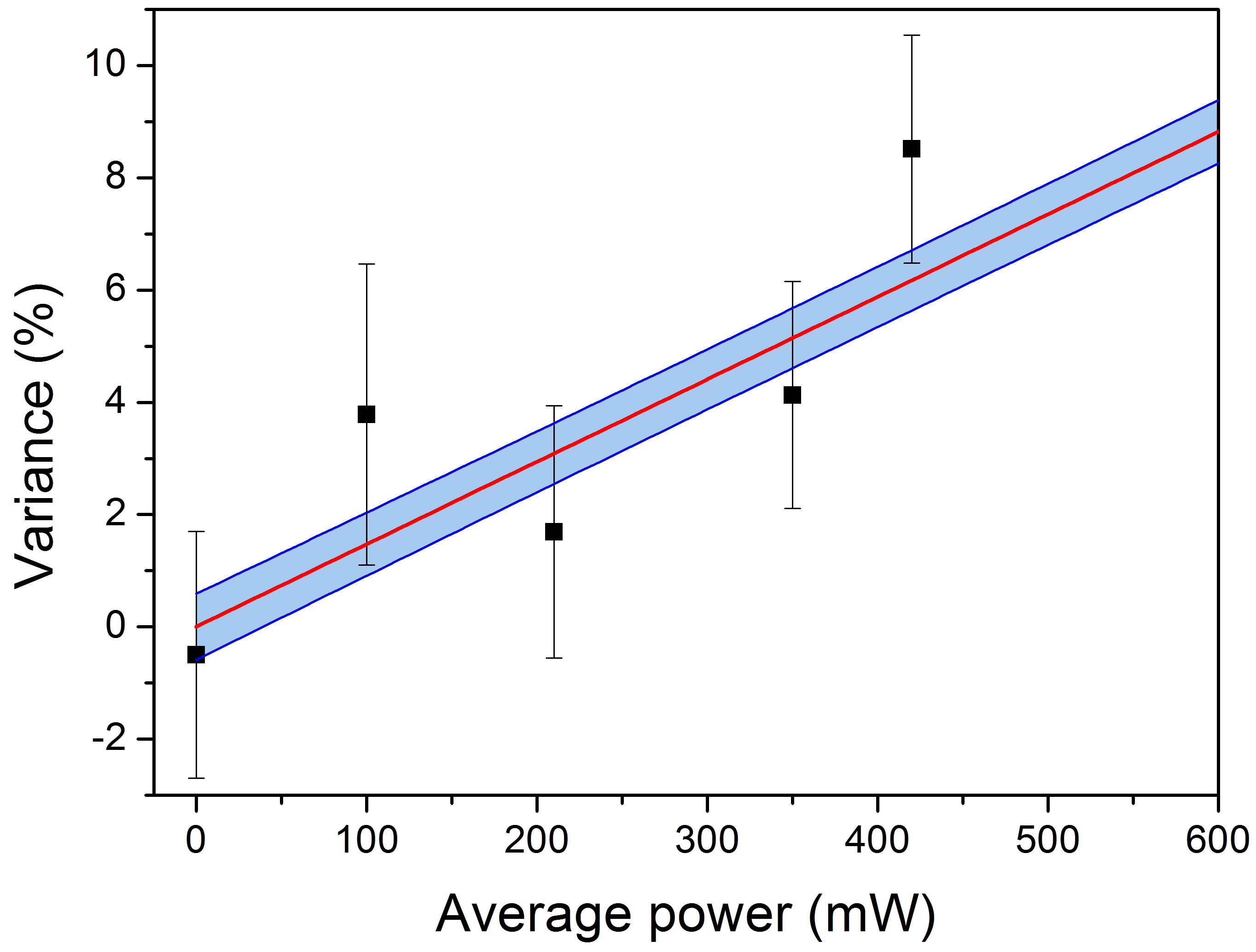}}
\caption{Experimental results of $C_{60}$ transverse manipulation, with standard error from
52 averaged measurement runs. (a) shows the number of molecules
detected as affected by switching the laser on (350 mW average
power) and off. (b) Average laser power dependence of the number
of molecules detected, the red line shows the results of Monte
Carlo simulations for a beam waist of 154 ${\mu}$m, the blue area
shows the 95$\%$ confidence range.}
\label{fig:results} \end{figure}

\emph{On-Off switching effect:} Fig.$\ref{fig:results}$(a) shows experimental
data of a series of nine consecutive measurements with the laser on or off. The average 'laser-on' power was 350 mW. Every data point represents the average of molecule counts over 13 minutes. We used a 0.5 mm pinhole in front of detector to measure only the center region of the molecular beam, where we expect an increase of detected molecules. We observe a clear modulation of the number of molecules being detected. Error bars are the standard deviation. Fluctuations of the detected signal are caused by molecular beam flux variations, laser instabilities as well as fluctuations in the QMS detector. The laser intensity was checked to be sufficiently stable for the time of the measurement. Integration time was chosen to reduce long term fluctuations while allowing for optimal signal to noise ratio from averaging. The experiment has been repeated several times with apertures of different size and shape in front of the detector and with a different molecule: tetra-phenyl porphyrin (TPP, 614 amu). All measurements support our observation of a transverse modulation of molecular motion. Although we observe only a small effect, this is the first experimental evidence for an optical dipole force effect on the center of mass motion of large molecules resulting from interactions with multiple light pulses. The spatial resolution of a scanning aperture method was not sufficient to image a focusing effect in the total beam profile. 

\emph{Linear power dependency:} To investigate the effect
further we vary the laser power and observe the number of
molecules detected. We observe a linear power dependency of the count rate in agreement with  Eqn.$~\ref{eq:vel}$ (see Fig.$~\ref{fig:results}$(b)). Data are an average of 52 measurement sequences taken over 15 seconds for each laser power subsequently, to reduce the effect of systematic count rate drifts. An maximal $8\%$ increase in total count rate was
observed for a maximum average laser power of 420 $mW$. This value is
replicated with our Monte Carlo simulations which are shown by the
red line in Fig.~\ref{fig:results}(b). Simulated trajectories of 10$^5$ molecules for different laser powers show the same linear dependency of the total molecule counts, in perfect agreement with the experiment for a laser beam waist of
$w_{0}$ = 154 ${\mu}$m, which was the only free parameter in the Monte Carlo simulations. This is in agreement with the optics setup of the experiment. Each molecule interacts on average with 63 light pulses. The maximum laser intensity at the center of the beam
waist is $4.4{\times}10^{8}$ $W/cm^{2}$.

\emph{Competing effects:} Arguably, a single 800 nm photon cannot ionize $C_{60}$. Multiphoton ionization becomes significant for intensities of approximately
$10^{13}W/cm^{2}$~\cite{Hunsche96}. In our experiment, the peak
intensity of pulses is of the order
${10}^{8}W/cm^{2}$, well below the ionization threshold.
It has been shown that femtosecond lasers can be used
to increase ionization rates in large molecules compared to
nanosecond pulses and to study the dynamics of the ionization process~\cite{Weinkauf94}.  In our experiment the average number of absorbed photons is:
$N_{abs}=\frac{2P\sigma \tau}{\pi \omega_0^2 h\nu}$, where $\tau$
is the interaction time between light of frequency $\nu$ and
molecule, $h$ is Planck's constant. We use the absorption
cross section of $\sigma=6{\times}10^{-20}$cm$^{-2}$ 
for $C_{60}$ at 800 nm~\cite{gotsche2007} to estimate an total average
value of $2.5{\times}10^{-4}$ photons absorbed by a molecule if it
passes through the center of the laser. With this we can exclude all
competing effects which depend on photo-absorption such as ionization,
fragmentation or dissociation to explain our observation. Furthermore the significance of photon
recoil effecting the center of mass motion of molecules can
be neglected~\cite{nimmrichter2008}.

We now argue that this manipulation technique is universal and applicable to any polarizable particle as both mass and polarizability scale with the volume of the particle. The polarizability to mass ratio for $C_{60}$ is given by $\alpha/$m = 0.1 ${\AA}^3$/amu. This ratio typically differs only by maximally $\pm 15\%$ for other molecules and particles~\cite{Bonin1997} and it is easily possible to change the optical potential $U$ by a factor of two through modulation of laser power, which would more than compensates the $\alpha$/m variation. The multiple pulse interaction may open the door to new light-molecule manipulation schemes as adding a new degree of freedom for handling.

In summary, we have observed a clear effect of multiple light pulses on the center of mass motion of neutral molecules. Further experiments are needed to optimize the light-molecule interaction effect. Simulations predict large deflection for high laser pulse energy as from ns-pulsed lasers, which have lower laser pulse repetition rates. Generally, the experiment can also be performed with high intensity continuous lasers. However more damage to the molecule is expected.

Acknowledgement: We thank the UK STFC laser loan pool for lending the laser, the UK South-East Physics Network (SEPnet) for a scholarship (P V), as well as the Foundational Questions Institute (FQXi) and the John F Templeton foundation for generous support.

\bibliography{refmole}
\end{document}